\documentclass[a4paper,12pt]{article}
\usepackage[left=20mm, right=2mm, bottom=20mm, top=30mm]{geometry}
\usepackage[english]{babel}
\usepackage[utf8]{inputenc}
\usepackage[T1]{fontenc}

\usepackage{mathptmx} % - Times New Roman font equivalent (required by EPJC)
\usepackage{amsmath}
\usepackage{epsfig}           % Encapsulated PostScript figure inclusion
\usepackage{xspace}
\usepackage{multirow}
\usepackage{enumerate}
\usepackage{flushend}
\usepackage[numbers,sort&compress]{natbib}
\usepackage{graphicx}
\usepackage{float}
\usepackage{wrapfig}
\usepackage{amssymb}
\usepackage{caption}
\usepackage{placeins}
\usepackage[colorlinks,citecolor=blue,urlcolor=blue,linkcolor=blue]{hyperref}

\begin{document}

To be submitted to Eur. Phys. J. C

\begin{center}
\section*{ Study of 
\fontfamily{ptm}\fontseries{b}\selectfont $K^+ \rightarrow \pi^{0} e^{+} \nu \gamma $ 
decay with OKA setup}
%\end{center}

\begin{minipage}{0.9\linewidth}
 \center{
  %\small 
  %\ttfamily
  %\rmfamily
  %\textsf
\center{
  %\small
 \rmfamily  A.Yu.~Polyarush\footnote{e-mail: polyarush@inr.ru},
  V.A.~Duk\footnote{\scriptsize Now at INFN-Sezione di Perugia, Via A. Pascoli, 06123 Perugia,Italy}, 
  S.N.~Filippov, E.N.~Guschin, 
  A.A.~Khudyakov, V.I.~Kravtsov, 
  Yu.G.~Kudenko\footnote{\scriptsize Also at Moscow Institute of Physics and Technology, Moscow, Russia}\footnote{\scriptsize Also at NRNU Moscow Engineering Physics Institute (MEPhI), Moscow, Russia}, 
}\vspace{-4mm}
 \center{\small 
    \textsc{(INR RAS, Moscow, Russia),}
 }\\

  \textsc
   S.A.~Akimenko, A.V.~Artamonov, 
  A.M.~Blik, V.S.~Burtovoy, S.V.~Donskov, A.P.~Filin, A.V.~Inyakin,
  A.M.~Gorin, G.V.~Khaustov, S.A.~Kholodenko,
  V.N.~Kolosov, V.F.~Kurshetsov, V.A.~Lishin,  M.V.~Medynsky,
  Yu.V.~Mikhailov, V.F.~Obraztsov, V.A.~Polyakov,  V.I.~Romanovsky,
  V.I.~Rykalin,  A.S.~Sadovsky, V.D.~Samoilenko, M.M.~Shapkin,
  O.V.~Stenyakin,  O.G.~Tchikilev, V.A.~Uvarov, O.P.~Yushchenko
 }
\vspace{-4mm}
 \center{\small 
   \textsc{(NRC "Kurchatov Institute"${}^{}_{}{}^{}$-${}^{}_{}{}^{}$IHEP, Protvino, Russia),} 
 }\\

 \vspace{-4mm}
 \center{
  \rmfamily
  V.N.~Bychkov, G.D.~Kekelidze, V.M.~Lysan, B.Zh.~Zalikhanov
 }\vspace{-4mm}
 \center{\small 
   \itshape 
   \textsc{(JINR, Dubna, Russia)}\\
 }
\end{minipage}
\end{center}

\vspace{5mm}
\abstract{
Results of a study of the 
$K^+ \rightarrow \pi^{0} e^{+} \nu \gamma $ decay
at OKA setup are presented. More than
32000 events of this decay are observed.
The differential spectra over the photon energy and the photon-electron
opening angle in kaon rest frame are presented. The branching ratios, normalized to that of 
$K_{e3}$ decay are calculated  for different cuts in $E^*_\gamma$ and $cos\Theta^{*}_{e\gamma}$.
In particular, the branching ratio for 
$E^{*}_{\gamma}>30$ MeV and 
$\Theta^{*}_{e \gamma}>20^{\circ}$ is measured 
R = $\frac{Br(K^+ \rightarrow \pi^{0} e^{+} \nu_{e} \gamma) }
{Br(K^+ \rightarrow \pi^{0} e^{+} \nu_{e})} $ = 
 (0.587$\pm$0.010($stat.$)$\pm$0.015($syst.$))$\times10^{-2}$,
which is in a good agreement with ChPT $O(p^{4})$ calculations.
}

\section{Introduction}
The decay 
$K^+ \rightarrow \pi^{0} e^{+} \nu \gamma $
provides fertile testing ground  for the Chiral Perturbation Theory (ChPT)
\cite{v1,v2}, the effective field theory of the Standard Model at low energies.

$K^+ \rightarrow \pi^{0} e^{+} \nu \gamma $ decay was first considered in \cite{v3}
up to the order ChPT $O(p^{4})$  and
branching ratios were  evaluated  for given cuts in the photon energy
and in the photon-electron opening angle in the kaon rest frame: $E^{*}_{\gamma} >E^{cut}_{\gamma}$, 
$\Theta^{*}_{e\gamma} > \Theta^{cut}_{e\gamma}$.
Later the CHPT analysis was revisited and extended to  $O(p^{6})$ \cite{v4}. The branchings at tree level were also calculated in papers \cite{v5,v6}, as well as T-odd correlations.

The matrix element for  
$K^+ \rightarrow \pi^{0} e^{+} \nu \gamma $ decay has general structure

$T =\frac{G_{F}}{\sqrt{2}}{e}V_{us}\epsilon^{\mu} 
(q)\Biggl\{(V_{\mu\nu}
- A_{\mu\nu})\overline{u}(p_{\nu})\gamma^{\nu}(1 - \gamma_{5})v(p_{e})\nonumber\\
+\frac{F_{\nu}}{2p_{e}q}\overline{u}(p_{\nu})\gamma^{\nu}(1 - 
\gamma_{5})(m_{e}-\hat{{p}_{e}}-\hat{q})\gamma_{\mu}v(p_{e})\Biggr\}\equiv
{\epsilon^{\mu}} A_{\mu}.$ \\
First term of the matrix element  describes the bremsstrahlung
 of kaon and the direct emission. The relevant diagram is displayed in Fig.1a. The 
lepton bremsstrahlung is presented by the second part of  Eq.(1) and Fig.1b.
The hadronic tensors $V^{had}_{\mu\nu}$ and $A^{had}_{\mu\nu}$ are defined by
$I_{\mu\nu} = i \int d^4 e^{iqx}\langle \pi^{0}(p')|TV^{em}_{\mu}(x)I^{had}_{\nu}(0)|K^+(p) \rangle$,\\ 
I = V, A, with
$V^{had}_{\nu} =  \overline s \gamma_{\nu} u$, 
$A^{had}_{\nu} =  \overline s \gamma_{\nu} \gamma_{5} u$,
$V^{em}_{\mu} = (2\overline u    \gamma_{\mu} u -  \overline d \gamma_{\mu} d - \overline s \gamma_{\mu} s   )/3$
and $F_{\nu}$ is the $K^+_{e3}$ matrix element
$F_{\nu}$ = $\langle \pi^{0}(p')|V^{had}_{\nu}(0)|K^+(p) \rangle$.

The bremsstrahlung part of the amplitude is largely dominant in the partial decay width.
Only with the advent of high statistics kaon decay experiments it become feasible to 
study effects of structure dependent contributions and  of the chiral anomaly. 
% fig 1
\begin{figure}[ht]
\hspace{0.5cm}
\mbox{
\includegraphics[scale=0.85]{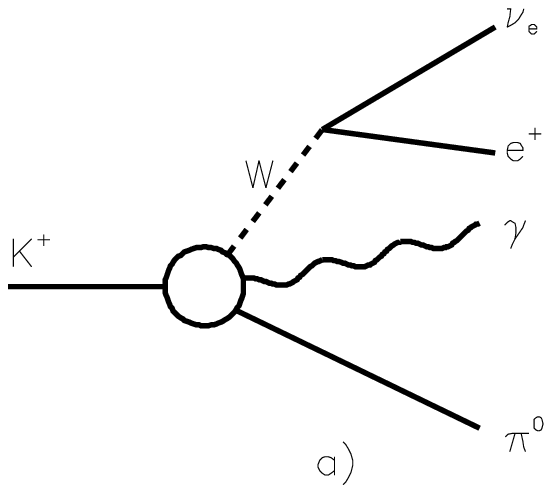}
\hspace{-1.0cm}
\includegraphics[scale=0.85]{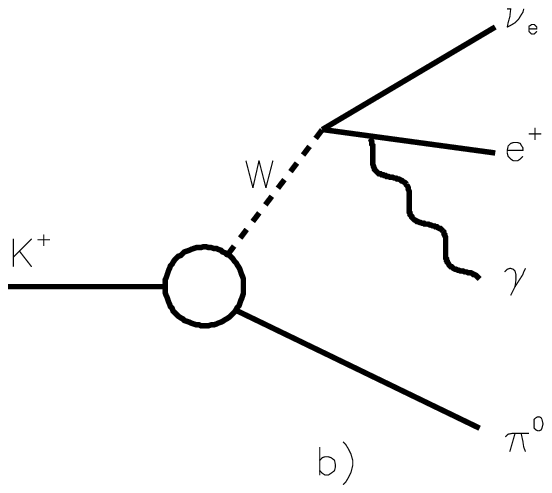}
}
\vspace{-1.5cm}
  \caption{Diagrams describing  $K^+ \rightarrow \pi^{0} e^{+} \nu \gamma $ decay }
%  \label{fig:piz59}
\end{figure} 

The numerical results given in \cite{v3,v4,v5,v6} demonstrate that non-trivial
CHPT effects can be detected by the experiment. 
This gives a motivation for the present study.

\section{ OKA setup } 
OKA collaboration operates at IHEP Protvino U-70 Proton Synchrotron.
OKA  detector (see Fig.2) is located in positive RF-separated beam with 12.5\% of kaon
with a momentum of 17.7 GeV/c and an intensity of $3{\cdot} 10^{5}$ kaons 
per 2 sec U-70 spill.
%Fig2.
\begin{figure}[!ht]
\center
 \includegraphics[width=0.9\textwidth]{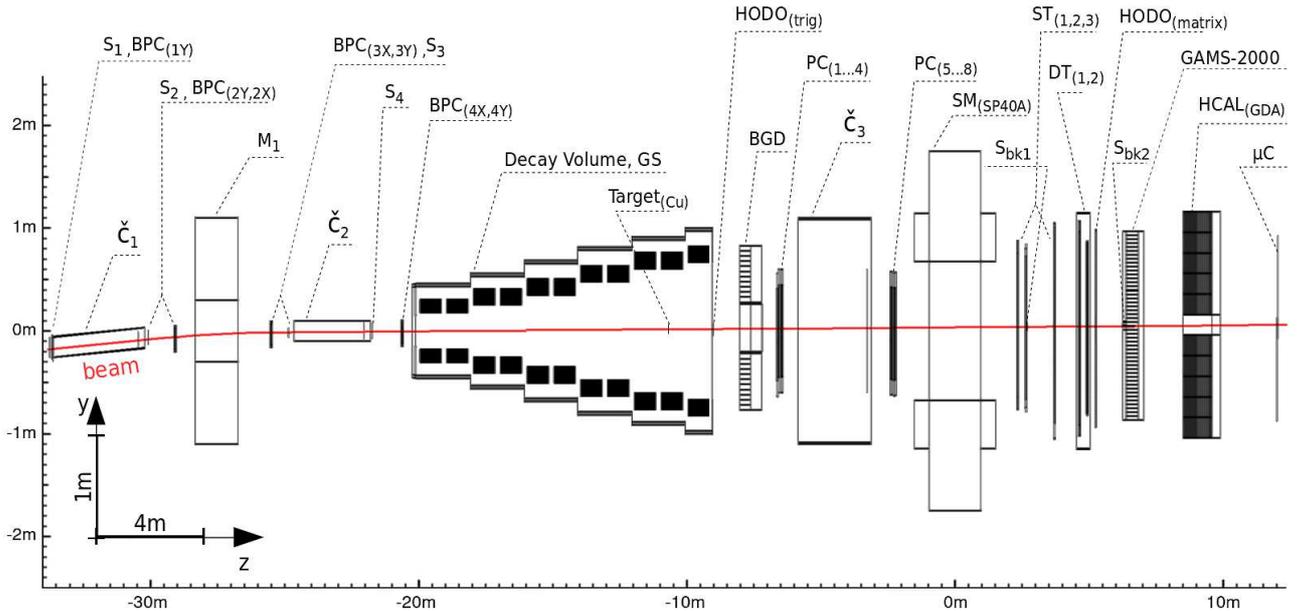}
  \caption{Layout of the OKA detector}

%  \label{fig:ust}
\end{figure}
RF-separation  with the Panofsky scheme is realized. 
It uses two superconductive Karlsruhe-CERN SC RF deflectors\cite{oka1}, 
donated by CERN. Sophisticated cryogenic system,
built at IHEP\cite{oka2} provides superfluid He for cavities cooling.
The detailed description of the OKA  detector
is given in our previous publications \cite{oka3,oka4}.
The OKA is taking data since 2010. 
In this study, we use  the statistics that was obtained 
in the 2012 and 2013 years.
The total  number of kaons 
entering the Decay volume (DV) corresponds to $\sim 3.4\times 10^{10}$.

Ruther simple trigger was used during data-taking:
${\tt Tr = S_{1}{\cdot}S_{2}{\cdot}S_{3}{\cdot}S_{4}{\cdot}\overline{\check{C}}_{1}
{\cdot}\check{C}_{2}{\cdot}\overline{S}_{bk}} {\cdot} (E_{GAMS} > 2.5GeV)$.
$S_{1}-S_{4}$  are scintillating counters; 
$\check{C}_{1}$, 
$\check{C}_{2}$ - Cherenkov counters (
$\check{C}_{1}$  sees pions, 
$\check{C}_{2}$  pions and kaons); 
$S_{bk}$ - two scintillation counters on the beam
axis after the magnet to suppress undecayed particles.

The MC simulation of the OKA setup is done within the GEANT3 framework\cite{v7}.
Signal and background events are weighted according to corresponding matrix elements.

\section{Events selection and background suppression}

Criteria for events selection:

1) One positive charged track detected in the tracking system and 4 showers detected in the electromagnetic 
calorimeters GAMS- 2000  and BGD. 

2) One shower must be associated with the charged track. 

3) The charged track is identified as a positron. The positron identification is done using the ratio of
the energy of the shower in GAMS- 2000 to the momentum of the
associated track. The $E/p$ distribution is shown in Fig.3. 
The particles with $0.8 < E/p < 1.2$ are accepted as positrons.
Another cut used for the suppression of the $\pi^{+}$ contamination  
is that    on the distance between the charged track extrapolation to the
front plane of the electromagnetic detector and the nearest shower.
This distance must be less than 3 cm.

% fig 3
%\begin{figure}[h!]
\begin{figure}[ht!]
\centering
\includegraphics[scale=0.85]{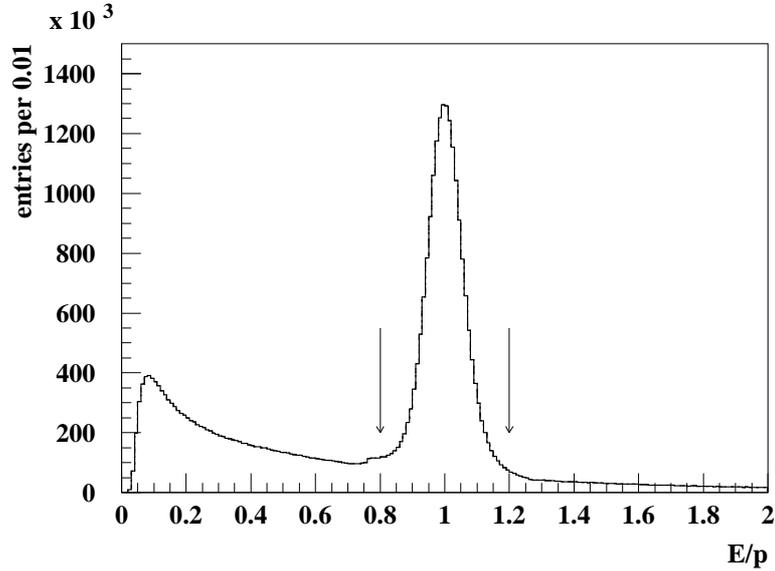}
\caption{E/p ratio for the real data.}
%\label{fig:piz9}
\end{figure}

4) The decay vertex situated within the decay volume.

5) The mass $M_{\gamma \gamma}$ of the $\gamma \gamma$ -- pair 
closest to the Table value of $\pi^{0}$ is  $0.12 < M_{\gamma\gamma} < 0.15$ GeV.
Energy of a photon originated from $\pi^{0}$ is greater than 0.5 GeV.
Energy of the radiative photon is greater than 0.8 GeV.
Absence of signals in veto system above noise threshold is also required.

The main background decay channels 
for the decay $K^+$$ \rightarrow $$\pi^{0} e^{+} \nu_{e}$$ \gamma $  are:

1) $K^+ \rightarrow \pi^{0} e^{+} \nu$ with an extra photon.     
The main source of  extra photons is the positron interactions in 
the detector.

2) $K^+ \rightarrow \pi^{+} \pi^{0} \pi^{0} $  where one  
of the $\pi^{0}$ photons is not detected  and $\pi^{+}$
is misidentified  as a positron.

3) $K^+ \rightarrow \pi^{+} \pi^{0}  $ with a ``fake  photon''   
and $\pi^{+}$ miss-identified as a positron. 
The fake photon clusters can come from $\pi n$ interaction in 
the gamma detector, and from accidentals.

4) $K^+ \rightarrow \pi^{+} \pi^{0} \gamma$  when $\pi^{+}$
is miss-identified as a positron.

5) $K^+ \rightarrow \pi^{0} \pi^{0}  e^{+} \nu$  when one $\gamma$ is lost.\\
All these background sources are included in our MC calculations.

To suppress the background channels we use a set of cuts:

Cut 1:  $E_{miss} > 0.5$ GeV.
The requirement on the missing energy mainly reduces  background (4).

Cut 2:   $\Delta y = | y_{\gamma}  - y_{e} | >$ 5 cm, where y is the vertical coordinate
of a particle in the electromagnetic calorimeter. (the magnetic field turns
charged particles in xz-plane).

Cut 3: $\mid x_{\nu},y_{\nu} \mid <$ 100cm.
The reconstructed missing momentum direction  must cross the active area
 of the electromagnetic calorimeter. 

Cut 4: $M_{K\rightarrow \pi^{0}e^{+}\nu_{e}\gamma} > 0.45$GeV.
$M_{K\rightarrow \pi^{0}e^{+}\nu_{e}\gamma}   $   - the reconstructed
mass of the ($\pi^{0} e^{+} \nu_{e}$$ \gamma $)- system, assuming $m_{\nu}=0$.
To inforce this cut   we use a cut on the missing mass squared 
$M^{2}$$(\pi^{0} e \gamma)$ = $(P_{K} - 
P_{\pi^{0}} - P_{e} - P_{\gamma})^2$. 
For the signal events this variable corresponds to the square of the neutrino 
mass and must be zero within measurement accuracy.

Cut 5: \hspace{0.8cm} $-0.003 < M^{2}(\pi^{0} e^{+} \gamma) < 0.003.$
The dominant background to $K_{e3\gamma}$ arises from 
$K_{e3}$ with an extra photon (background (1)). 
This background is suppressed by a requirement on the
 angle between positron and photon
in the laboratory frame  $\Theta_{e \gamma}$  (see Fig.4).
The distribution of the $K_{e3}$-background events
has a very sharp peak at zero angle. This peak is significantly 
narrower than that for the signal events. This happens, in particular, 
because the emission of the photons by the positron 
occurs in the setup material downstream the decay vertex,
but the angle is still calculated as if emission comes from the vertex.

Cut 6: \hspace{0.8cm} $0.004 < \Theta_{e \gamma}$ $ <0.080.$ 
%\vspace{0.3cm}
Left part of this cut is introduced exactly for the suppression of background (1).
The right cut is against $K_{\pi2}$ background.

\begin{figure}[ht]
\hspace{0.0cm}
\centering
\includegraphics[scale=0.85]{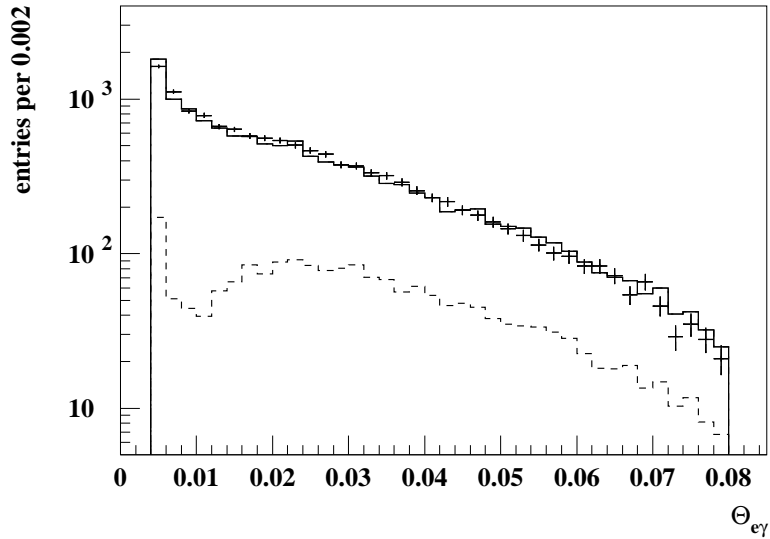}
\caption{ Distribution    over  $\Theta_{e \gamma}$ - the angle between
 positron and photon in lab. system. Real data  (points with errors), 
signal plus MC background  (solid line histogram),
MC background (dotted line histogram). }
%\label{fig:piz9}
\end{figure}

After all the cuts, 
32676 candidates are selected, with a background of 
4624 events.
Background normalization is done by comparison of the number of events for  
$K_{e3}$ decay in MC and real data samples.

\section {Results }
The resulting distribution of the selected events over 
cos$(\Theta^{*}_{e\gamma})$, $\Theta^{*}_{e\gamma}$ being the
angle between the positron and the photon in
the kaon rest frame, is shown in Fig.5. 
% fig 5/
\begin{figure}[!ht]
\centering
\includegraphics[scale=0.85]{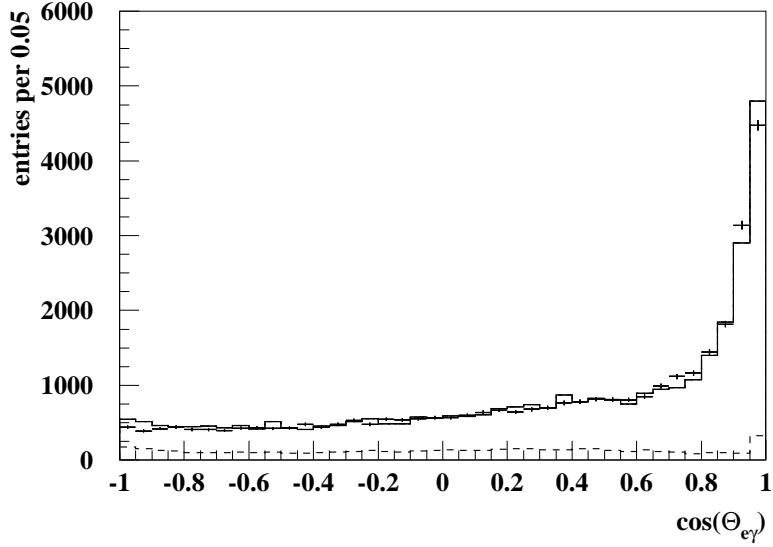}
\caption{ 
The distribution of the events over $cos\Theta^{*}_{e\gamma}$.
Points with errors -  the real data, histogram -  MC signal plus background,
MC background - dotted line histogram. }

%\label{fig:piz10}
\end{figure}

The distribution over  $E^{*}_{e\gamma}$ - the photon energy in the kaon rest
frame is shown in Fig.6.  Reasonable agreement of the
data with MC is seen. When generating the signal MC, a generator 
based on  $O(p^4)$ calculations\cite{v3} is used. 

\begin{figure}[ht]
\centering

\includegraphics[scale=0.85]{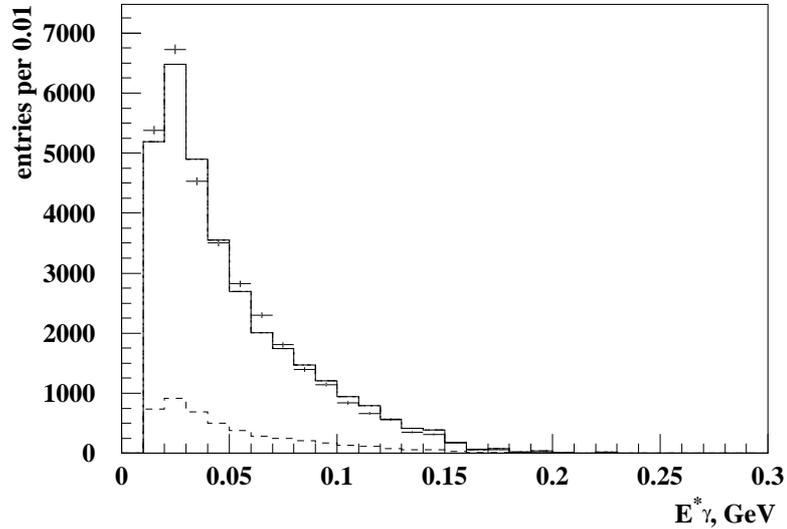}
\caption{The distribution of the events over $E^{*}_{e\gamma}$.
Points with errors is the real data, histogram - MC signal plus background,
MC background - dotted line histogram.}
%\label{fig:piz11}
\end{figure}

To obtain the branching ratio for the $K_{\pi^{0} e^{+}\nu_{e}\gamma}$ relative 
to the  $K_{e3}$ (R), 
the background and efficiency corrected number of
$K_{e3\gamma}$ events is normalized on that of about 9M $K_{e3}$ events
found with a similar selection criteria. 

The relative branching ratio (R) for the soft cuts  $E^{*}_{\gamma} > 10$ MeV 
and $\Theta^{*}_{e\gamma} > 10^{\circ}$ is found to be

  $R_1= \frac{Br(K^+ \rightarrow \pi^{0} e^{+} \nu_{e} \gamma) }
{Br(K^+ \rightarrow \pi^{0} e^{+} \nu_{e})} = 
  (1.990\pm0.017(stat.)\pm0.021(syst.))\times10^{-2}$.
\\
And for the cuts  $E^{*}_{\gamma} > 30$ MeV and $\Theta^{*}_{e\gamma} > 20^{\circ}$ 
used for the comparison with theory we have

  $R_2= \frac{Br(K^+ \rightarrow \pi^{0} e^{+} \nu_{e} \gamma) }
{Br(K^+ \rightarrow \pi^{0} e^{+} \nu_{e})} = 
 (0.587\pm 0.010(stat.) \pm 0.015(syst.)) \times10^{-2}$.\\
For the comparison with previous experiments the branching ratio
with the cuts $E^{*}_{\gamma}>10$ MeV, 
$0.6<cos\Theta^{*}_{e \gamma}<0.9$ is calculated
   
$R_3= \frac{Br(K^+ \rightarrow \pi^{0} e^{+} \nu_{e} \gamma) }
{Br(K^+ \rightarrow \pi^{0} e^{+} \nu_{e})} = 
(0.532\pm0.010(stat.)\pm0.012(syst.))\times10^{-2}$.

Systematic errors are estimated by variation of the cuts 1-6.
Contributions of each cut variation to systematic errors are given in  Table 1.

\begin{table}
\caption{Contributions to systematic errors. }   
\begin{center}
\begin{tabular}{ccccccc}
\hline
$R_{i}$ & 1 &2 &3 & 4  & 5 & 6 \\
\hline
$ R_1 $& 0.003 & 0.002   & 0.006 &  0.008 &  0.015 &  0.011 \\
$ R_2 $& 0.004 & 0.001   & 0.004 &  0.005 &  0.010 &  0.008 \\
$ R_3 $& 0.001 & 0.001   & 0.005 &  0.001 &  0.010 &  0.004 \\

\hline
\end{tabular}
\end{center}
\end{table}

The comparison with previous experiments is given in  Table 2.

\begin{table}[ht]
\caption{Br$(K^+ \rightarrow \pi^{0} e^{+} \nu_{e} \gamma)$/Br$(K^+ \rightarrow \pi^{0} e^{+} \nu_{e})$ for 
 $E^{*}_{\gamma} > 10$ MeV, $0.6<cos\Theta^{*}_{e\gamma}<0.9$  in comparison with  previous data. }   
\begin{center}
\begin{tabular}{lrr}
\hline
$R_{3}\times 10^{2}$ & $N_{ev}$ & experiment   \\
\hline
$ 0.53\pm0.01\pm0.01 $&  7248  & this experiment    \\
$ 0.48\pm0.02\pm0.03 $&  1423  & ISTRA+  \cite{e3}  \\
$ 0.46\pm0.08 $&    82  & XEBC \cite{e4}     \\
$ 0.56\pm0.04 $&   192  & ISTRA \cite{e5}     \\
$ 0.76\pm0.28 $&    13  & HLBC \cite{e6}     \\
\hline
\end{tabular}
\end{center}
\end{table}
\FloatBarrier
\section*{Conclusions}
The largest statistics of about 32K events of $K_{e3\gamma}$  is collected by the OKA experiment. 
The relative branching ratio R=Br$(K^+ \rightarrow \pi^{0} e^{+} \nu_{e} \gamma)$/Br$(K^+ \rightarrow \pi^{0} e^{+} \nu_{e})$
 is measured
for different cuts on the photon energy and the photon-electron angle in the kaon rest frame. 
The obtained value of R for $E^{*}_{\gamma} > 30$ MeV and $\Theta^{*}_{e\gamma} > 20^{\circ}$
is in a good agreement with the CHPT $O(p^4)$ prediction\cite{v3}
R=(0.592$\pm$0.005)$\times10^{-2}$
and is some 2-3$\sigma$  away from the tree level results\cite{v5,v6}. 
The  $O(p^6)$ result\cite{v4} is ~2.5$\sigma$ higher.
That is, the measurement becomes sensitive to the non-trivial CHPT effects.

\section*{Acknowledgements}

The authors express their gratitude to the colleagues from the accelerator Department
for good performance of the U-70 during data taking; to colleagues from the beam department
for the stable operation of the 21K beam line, including RF-deflectors, and to colleagues
from the engineering physics department for the operation of the cryogenic system of the RF-deflectors.

The work is supported in part by the Russian Fund for Basic Research, grant N18-02-00179A.

\end{document}